\documentclass{pnastwo}

\usepackage[dvips]{graphicx}

\usepackage{amssymb,amsfonts,amsmath}

\usepackage{color}
\newcommand{\bqn}{\begin{eqnarray}}
\newcommand{\eqn}{\end{eqnarray}}
\newcommand{\bi}{\begin{itemize}}
\newcommand{\ei}{\end{itemize}}

\def\pomega{\varpi}

\def\icarus{{Icarus}}

\def\pomega{\varpi}
\def\deg{^\circ}
\newcommand{\y}{{}}
\newcommand{\apj}{{ApJ}}

\newcommand{\apjl}{{ApJLett}}
\newcommand{\mnras}{{MNRAS}}

\newcommand{\nat}{{Nature}}

\newcommand{\aap}{{A\&Ap}}

\newcommand{\araa}{{Annual Review of Astronomy \& Astrophysics}}
\newcommand{\pasp}{{PASP}}

\contributor{Submitted to Proceedings
of the National Academy of Sciences of the United States of America}
\url{www.pnas.org/cgi/doi/10.1073/pnas.0709640104}
\copyrightyear{2008}
\issuedate{Issue Date}
\volume{Volume}
\issuenumber{Issue Number}

\begin{document}



\title{Secular chaos and its application to Mercury, hot Jupiters, and the organization of planetary systems}

\author{Yoram Lithwick\affil{1}{Dept. of Physics and Astronomy, Northwestern University, 2145 Sheridan Rd., Evanston, IL 60208
\& Center for Interdisciplinary Exploration and Research in Astrophysics (CIERA)},
Yanqin Wu\affil{2}{Department of Astronomy and Astrophysics, University of Toronto, Toronto, ON M5S 3H4, Canada}}

\contributor{Submitted to Proceedings of the National Academy of Sciences
of the United States of America}

\maketitle

\begin{article}

\begin{abstract} 
  In the inner solar system, the planets' orbits evolve chaotically,
  driven primarily by secular chaos.  Mercury has a particularly
  chaotic orbit, and is in danger of being lost within a few billion
  years.  Just as secular chaos is reorganizing the solar system
  today, so it has likely helped organize it in the past.
   We suggest that 
 extrasolar planetary systems are
also organized to a large extent by secular chaos.  
 A hot Jupiter
 could be the end state
of   a secularly chaotic planetary system reminiscent of the
  solar system.  But in the case of the hot Jupiter, the innermost
  planet was Jupiter- (rather than Mercury-) sized, and its chaotic
  evolution was terminated when it was tidally captured by its star.
In this contribution, we review our recent work elucidating the physics of
secular chaos and applying it to Mercury and to hot Jupiters.  {\y We
  also present new results comparing} the inclinations of hot Jupiters
thus produced 
 with observations.
 \end{abstract}


\section{Introduction}

\begin{figure}
 \centerline{\includegraphics[width=.5\textwidth]{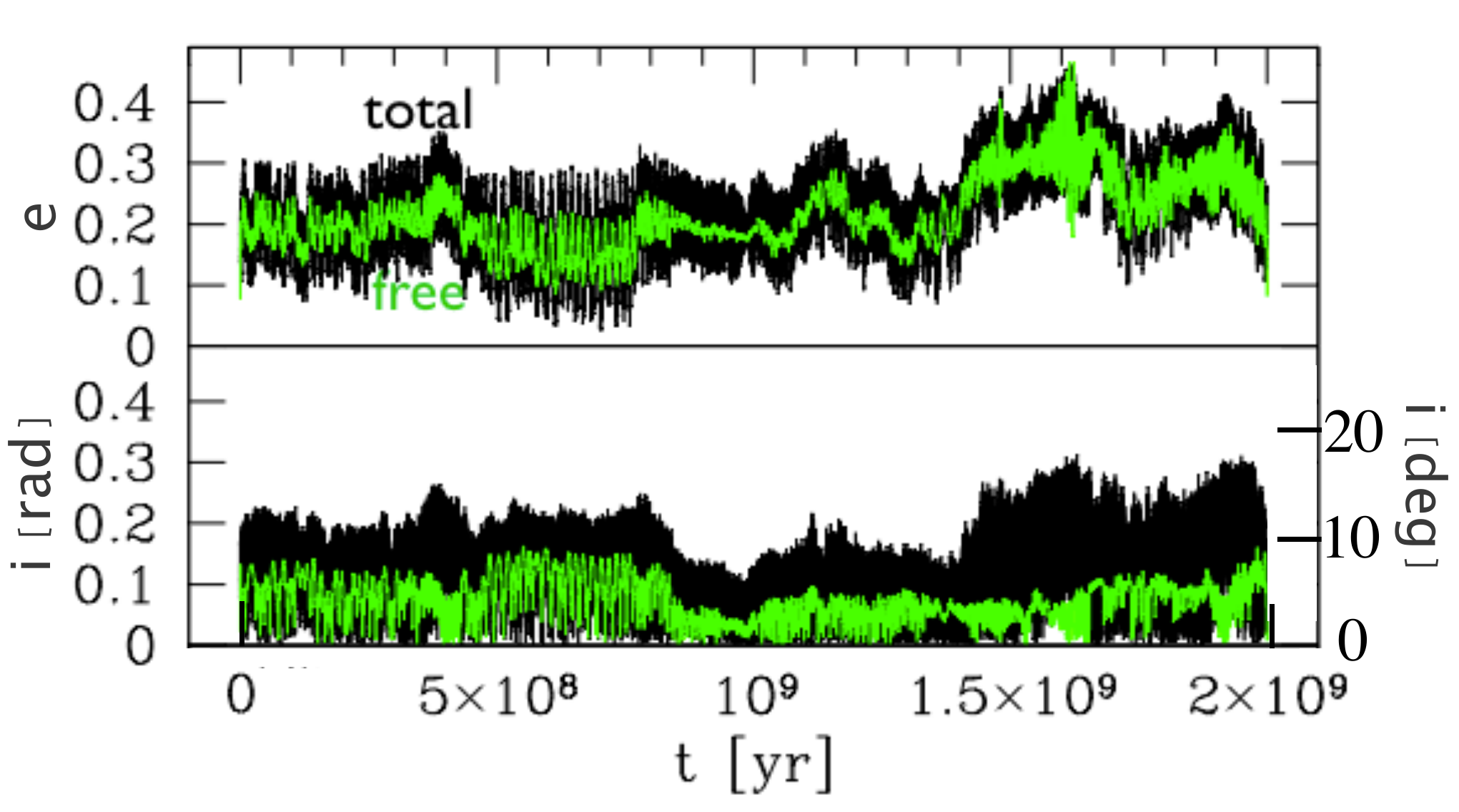}}
 \caption{Mercury's chaos in an N-body integration of the solar system.  
 Black  shows Mercury's total eccentricity and inclination.
 Green shows
  free $e$ and $i$, which were obtained
 by  filtering out the forcing frequencies of the other planets
  in Fourier space. 
   The wander of the green curves
  demonstrates that Mercury's orbit is chaotic.
     Adapted from  LW11.
 \label{fig:merc}}
 \end{figure}

\dropcap{T}he question of the stability of the solar system has a long
and illustrious history  (e.g., \cite{Laskar96}).  
{\y It was finally answered} with the aid of computer simulations
\cite{SussmanWisdom88, Laskar89, WisdomHolman}, which have shown that
the solar system is {\it marginally stable}: it is chaotically
unstable, but on a timescale comparable to its age.  In the inner
solar system, the planets' eccentricities {\y ($e$)} and inclinations
{\y ($i$)} diffuse in billions of years, with the two lightest
planets, Mercury (Fig. \ref{fig:merc}) and Mars, experiencing
particularly large variations.  Mercury may even be lost from the
solar system on a billion-year timescale
\cite{Laskar08,Batygin,LaskarGastineau}\footnote{\y Chaos is much
  weaker in the outer solar system  than in the
  inner  \cite{murrayholman99,Graz2005,2008MNRAS.386..295H}}.
Yet despite the spectacular success in solving solar system stability,
fundamental questions remain: What 
{\y is the theoretical explanation for} orbital chaos of the solar
system?
What does the chaotic nature of
the solar system teach us about its history and organization?  And how
does this relate to extrasolar systems?

For well-spaced planets  that are not close to strong 
  mean-motion-resonances (MMR's), 
  the orbits
evolve on timescales much longer than orbital periods.  Hence one may
often 
simplify the problem by
orbit-averaging the interplanetary interactions. The averaged
equations are known as the {\it secular} equations
(e.g., \cite{MD00}).
{\y To linear order in masses, secular} 
evolution consists of interactions between rings, which 
{\y represent the planets after their masses have been smeared out
  over an orbit.}
 Secular evolution
dominates the evolution of the terrestrial planets in the solar system 
{\y \cite{Laskar08}}, and it is natural
to suppose that it dominates in many extrasolar systems as well.  
  This is the type of planetary interaction we focus on in this
  contribution.

 One might be tempted by the  small eccentricities
and inclinations in the  solar system to simplify 
further
 and linearize the secular equations, i.e., 
 consider only terms to leading order in eccentricity and inclination.  
 Linear secular theory 
 reduces
 to a simple eigenvalue problem.  For $N$ secularly interacting
 planets, the solution consists of $2N$ eigenmodes: $N$ for the
 eccentricity degrees of freedom and another $N$ for the
 inclination. Each of the eigenmodes evolves independently of the
 others \cite{MD00}.  Linear secular theory provides a satisfactory
 description of the planets' orbits on million year
 timescales.\footnote{Higher accuracy can be achieved by including the
   most important MMR's
   \cite{MD00}.}  It is the cause, for example, of Earth's
 eccentricity-driven Milankovitch cycle.  But on timescales $\gtrsim
 10^7$ years, the evolution is chaotic (e.g., Fig. \ref{fig:merc}), in
 sharp contrast to the prediction of linear secular theory.  That
 appears to be puzzling, given the small eccentricities and
 inclinations in the solar system.

 {\y Yet despite its importance, there has been little theoretical
   understanding of how secular chaos works.  Conversely, chaos driven
   by MMR's is much better understood.  For example, chaos due to MMR
   overlap explains the Kirkwood gaps in the asteroid belt
   \cite{1983Icar...56...51W}, and chaos due to the overlap of 3-body
   MMR's accounts for the very weak chaos in the outer solar system
   \cite{murrayholman99}. Since   chaos in the
   solar system is typically driven by overlapping resonances (e.g., see review \cite{Lecar}),
    one might reason that  the secular chaos of the inner solar system is driven by
   overlapping {\it secular resonances}.  Laskar and Sussman \& Wisdom
   \cite{laskar90,laskar92,SussmanWisdom} 
    identified a number of candidate secular
   resonances that might drive chaos in the inner solar system by examining
   angle combinations that alternately librated and circulated in
   their simulations.  But there are an infinite number of such angle
   combinations, and it is not clear which are dynamically
   important---or why \cite{SussmanWisdom,Lecar}.}

 In \cite{lithmerc} (hereafter LW11) we
      {\y developed the theory for secular chaos,
   and applied it to Mercury, the solar system's most unstable planet.
   We   demonstrated how the locations and widths of
    general secular resonances  can be calculated, and
    how the overlap of the relevant
   resonances quantitatively explains Mercury's chaotic orbit.  This
   theory, which we review below, shows why Mercury's motion is
   nonlinear---and chaotic---even though $e$'s and $i$'s remain
   modest.  } {\y It also shows that Mercury lies just above the
   threshold for chaotic diffusion.}

A system of just two secularly interacting planets can be chaotic if
their eccentricities and inclinations are both of order unity
\cite{Libert,Migas09,Naoz}.  In systems with three or more planets,
on the other hand, there is a less stringent criterion on the minimum
eccentricity and inclination required for chaos, and the character of
secular chaos is more diffusive.  This diffusive type of secular chaos
promotes equipartition between different secular degrees of freedom.
During secular chaos, {\y the angular momentum of each planet varies
  chaotically, with the innermost planet being slightly more
  susceptible to large variations 
  (\cite{WuLithwick}, hereafter WL11).
   If enough angular momentum is removed from that
  planet, its pericenter will} 
approach the star.  
{\y And if that} planet 
{\y
  resembles  Jupiter}, tidal interaction with its host star may
then remove {\y its} orbital energy, turning it
into a hot Jupiter. Hot Saturns or hot Neptunes may also be produced
similarly. As shown in 
{\y WL11} and reviewed below, such a migration mechanism can reproduce
a range of observed features of hot Jupiters, giant planets that orbit
their host stars at periods of a few days.
It differs from other mechanisms that have been proposed for migrating hot Jupiters, including disk migration
\cite{Lin86,Lin96}, planet scattering \cite{FordRasio2}, and Kozai migration by a stellar or planetary
companion
\cite{WuMurray,FabryckyTremaine,Naoz}.

These studies
 prompt us to suggest
 that secular chaos may
play an important role in reshaping planetary systems after they
emerge from their nascent disks. 
     {\y Secular chaos  causes planets' eccentricities
    to randomly wander. 
     When one of the planets
attains high enough $e$ that it suffers collision, ejection, or tidal
capture,
the removal of that planet can then lead to a
more stable system, with a longer chaotic diffusion time.}  Such a
scenario (e.g., \cite{Laskar96}) can explain
why 
{\y the} solar system, as well as many observed exo-planetary systems,
are perched on the threshold of instability.

\section{Theory of Secular Chaos}

\subsection{Linear Secular Theory} 
    
We review  first linear secular theory before introducing nonlinear effects.
 The  equations of motion may be derived from the expression for the energy, which we shall label $H$ because it  turns into the Hamiltonian after replacing orbital elements with canonical
variables. Focusing  first on two coplanar planets,  their secular interaction energy is, to leading order in eccentricities and dropping constant terms, 
\begin{equation}
H = -G{mm'\over a'}\left( f_2(e^2+e'^2)+f_{10}ee'\cos\left({\pomega-\pomega'} \right)
\right) \ ,
\label{eq:hamlin}
\end{equation}
following the  notation of \cite{MD00} and dropping higher order terms
in $e$ (which lead to nonlinear equations). 
Here, 
 unprimed and primed quantities
denote the inner and outer planets and $f_j$ are Laplace coefficients that
are functions of $\alpha\equiv a/a'$ (Appendix B in \cite{MD00}).
In secular theory the semimajor axes are constant (even to nonlinear order), and hence may 
be considered as parameters. 
To derive the equations of motion for the inner planet, one replaces $e$ and $\pomega$ in the
interaction energy with
a canonically conjugate  pair (e.g., the Poincar\'e variables $\Gamma\approx m({GM_* a})^{1/2}e^2/2$ and $\gamma=-\pomega$) and employs the usual Hamilton's equations for $\Gamma$ and $\gamma$.  The same is true for the outer planet. One finds, after writing the resulting equations in terms of complex eccentricities ($z\equiv ee^{i\pomega}$ and $z'\equiv e'e^{i\pomega'}$)
\bqn
{d\over dt}\left(
\begin{array}{c} z\\z'\end{array}
\right)= 
i {2f_2n'\over \sqrt{a}}
\left(
\begin{array}{cc} \mu'\sqrt{a'}& -\beta\mu'\sqrt{a'}
\\
-\beta\mu\sqrt{a}& \mu\sqrt{a}
  \end{array}
\right)
\left(
\begin{array}{c} z\\z'\end{array}
\right)
\label{eq:linsec}
\eqn where $n'=\sqrt{GM_*/a'^3}$, $\mu=m/M_*$, and
$\beta=-f_{10}/(2f_2)$ 
{\y (For example, $\beta\approx 5\alpha/4$ for small $\alpha$.)}  The
solution to this linear set of equations is a sum of two eigenmodes,
each of which has a constant amplitude and a longitude that precesses
uniformly in time.  The theory may be trivially extended to $N$
planets, leading to $N$ eigenmodes.  It may also be extended to linear
order in inclinations, which leads to a second set of $N$ eigenmodes.
The equations for $\zeta=ie^{i\Omega}$ are identical to those for $z$,
but with $\beta\rightarrow 1$.

\subsection{Overlap of Secular Resonances Drives Secular Chaos}
The linear equations admit the possibility for a secular resonance, which occurs when 
two  eigenfrequencies match.
Consider a massless particle  perturbed by a precessing mode.
In anticipation of application to the solar system, one may think of the test particle
as Mercury, and the mode as the one dominated by Jupiter.
Equation (\ref{eq:linsec}) implies that  the particle's $z$ is governed by
\bqn
{dz\over dt}= i\gamma \left( z - e_m e^{ig_mt}\right) \ ,
\label{eq:zdotres}
\eqn
where $\gamma$ is the particle's free precession rate, 
$g_m$ is the mode's precession rate, and $e_m$ is proportional 
to the mode's amplitude (i.e., to the eccentricities of the massive planets that participate
in the mode). 
To order of magnitude,  $\gamma\sim n\mu_*\alpha_*^3$ and $e_m\sim\alpha_* e_*$, 
where starred quantities correspond to the planets that dominate the forcing
(and assuming $\alpha_*\ll 1$).
The solution to equation (\ref{eq:zdotres}) is a sum of free and forced eccentricities:
\bqn
z= {\rm const}\times e^{i\gamma t}+e_m{\gamma\over  \gamma-g_m} e^{i g_mt} \ .
\label{eq:zlin}
\eqn The free eccentricity exists even in the absence of the mode, and
it precess at frequency $\gamma$.  The forced eccentricity precesses
at the frequency of the mode that drives it, and its amplitude is
proportional to that mode's amplitude.  Formally, it diverges at
resonance, $\gamma =g_m$.  But nonlinearities alter that conclusion.

The leading nonlinear
correction to Hamiltonian (\ref{eq:hamlin}) is fourth order in eccentricity, 
which 
 changes Equation (\ref{eq:zdotres}) to\footnote {\y 
Eq. \ref{eq:znonlin} assumes
 $\alpha\ll 1$ (for algebraic simplicity).
 Many extrasolar systems have $\alpha\sim 1$, 
  \newline
  but
 that  only changes some numerical
 coefficients. }.
\bqn
{dz\over dt}=i\gamma\left(
z\left(1-{|z|^2\over 2}\right)
-e_m e^{ig_m t}\right) \
\label{eq:znonlin} 
\eqn
Hence nonlinearity reduces the frequency of free precession from
$\gamma$ to $\gamma(1-e^2/2)$.  There are a number of interesting
consequences.  First, even if the particle is at exact linear
resonance ($\gamma=g_m$), then as its eccentricity changes its
frequency shifts away from resonance, protecting it against the
divergence that appears in linear theory (eq. \ref{eq:zlin}).  Second,
if the particle is not at linear resonance it can still approach
resonance when its eccentricity changes.  With nonlinearity included,
a resonance takes on the familiar shape of a ``cat's eye'' in phase
space, and a particle can librate stably in resonance (Fig. 1 of LW11).

 Although the nonlinear cat's eye protects against divergences, danger
 lurks at the corner of a cat's eye: an unstable fixed point.  Motion
 due to a single resonance (e.g., Eq. \ref{eq:znonlin}) is perfectly
 regular (non-chaotic).  But if there is a second resonance nearby in
 phase space---in particular, if the separatrices enclosing two
 different cats' eyes overlap---chaos ensues (Fig 2 of LW11).
  Chaos due to the overlapping of resonances
 drives Mercury's long-term evolution, and {\y  may
   well be} one of the key drivers of the long-term evolution of
 planetary systems.

\section{Mercury}

\begin{figure}
 \centerline{\includegraphics[width=0.42\textwidth,trim=20 120 50 80,clip=true]{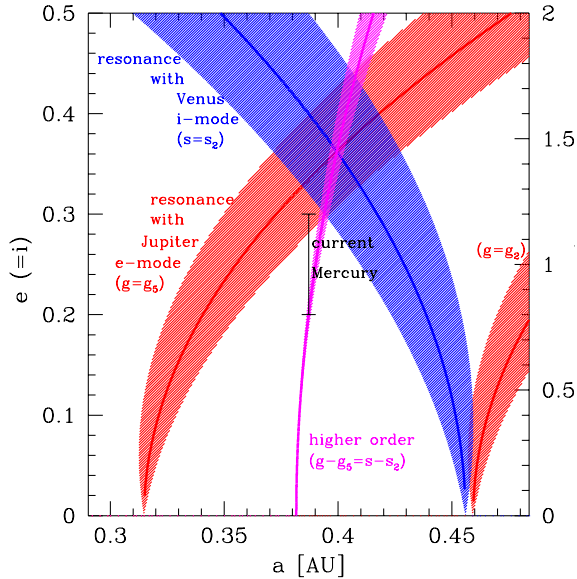}}
 \caption{Secular resonances that lead to Mercury's orbital chaos, {\y
      showing that Mercury's current orbit is close to the overlap of
     two secular resonances (i.e., $g=g_5$ with $s=s_2$)}. 
{\y To plot the}
   lines at the center of these resonances {\y we} assume the current
   values of $g_5$ and $s_2$, and that Mercury remains with $e=i$.
   The widths of the resonances, {\y as calculated in LW11,} 
are shaded.  The widths depend not
   only on Mercury's orbital parameters, but on the other planets' as
   well.  To plot these widths, we scale {\it all} of the planets'
   $e$'s and $i$'s by the same factor ($\kappa$; see right axis)
   relative to their current values.  We do this to illustrate how the
   solar system is perched on the threshold of chaos.  
 For
   this plot, the widths are only correct within a factor of $\lesssim
   2$ because the exact width {\y (see LW11)}  depends on the trajectory of an orbit in
   $e$-$i$ space.
   \label{fig:resoverlap}}
 \end{figure}

{\y The theory described above for coplanar secular chaos was first
  developed by \cite{sid90}.  But to explain secular chaos in the
  solar system, one must extend it to include non-zero inclinations,
  which we did in LW11. } Mercury has two free frequencies, one for
its eccentricity ($z$) and one for its inclination
($\zeta$)\footnote{We continue to treat Mercury as massless, which is
  an adequate approximation (LW11).}. We denote these
$g$ and $s$ respectively. In linear theory, $g=-s=\gamma$.  But to
leading nonlinear order, these frequencies are modified in the manner
described above to \bqn g&=&\gamma\left(1-{e^2\over 2}-2i^2
\right) \label{eq:g}
\\
s&=&\gamma\left(-1-2e^2+{i^2\over 2}\right) \label{eq:s} \eqn
(LW11).  Each of these frequencies can be in resonance
with one of the other 13 planetary modes---2 for each planet,
excluding the zero frequency inclination mode that defines the
invariable plane.\footnote{Another resonance---the Kozai
  resonance---occurs at high enough $i$ so that $g=s$, although
  \newline at such high $i$'s, our expansion to leading nonlinear
  order is suspect.} Two solar system modes have frequencies close to
Mercury's linear eccentricity precession rate (i.e., to $\gamma$): the
Venus-dominated $e$-mode (frequency $g_2$) and the Jupiter-dominated
$e$-mode ($g_5$).  And one mode's frequency lies close to Mercury's
linear inclination precession rate (i.e., to $-\gamma$): the
Venus-dominated $i$-mode ($s_2$).  One can visualize this by imagining
moving Mercury's $a$, holding its $e=i=0$.  Since $\gamma$ is a
function of $a$, the linear secular resonances occur at particular
values of $a$.  The three aforementioned resonances lie $\sim 20\%$
away from Mercury's actual $a$ at $e=i=0$ (Fig. \ref{fig:resoverlap}).
Since Mercury is at some distance from those resonances to linear
order, it is at first sight surprising that they can play an important
role in Mercury's evolution.  But Equations (\ref{eq:g})--(\ref{eq:s})
show that the locations of these resonances move as Mercury's $e$ and
$i$ are increased.  In fact, two of them ($g_5$ and $s_2$) overlap
very close to Mercury's current orbital parameters.  The overlapping
of those two resonances is the underlying cause of Mercury's chaos.
Even though Mercury has relatively small $e$ and $i$, its proximity to
two secular resonances drives its orbit to be chaotic.

To make the above theory more precise, one must  calculate the widths of the
resonances, which are sketched in Figure \ref{fig:resoverlap}. 
If the resonant widths
are negligibly small, Mercury would have to lie precisely at the region of overlap, 
which would be unlikely. 
One also has to 
account for higher order (combinatorial) resonances, the most important one of which
is $(g-g_5)-(s-s_2)$.  That combinatorial resonance was  identified  by {\y Laskar} \cite{laskar90} from the
fact that it  librated  in his simulations for 200 Myr.
In  LW11, we calculated the widths of  the aforementioned resonances, and showed
that the widths match in detail what is seen in simulations (see Figs. 4, 6, and 7 in that paper).  

\begin{figure}
  \centerline{\includegraphics[width=.45\textwidth]{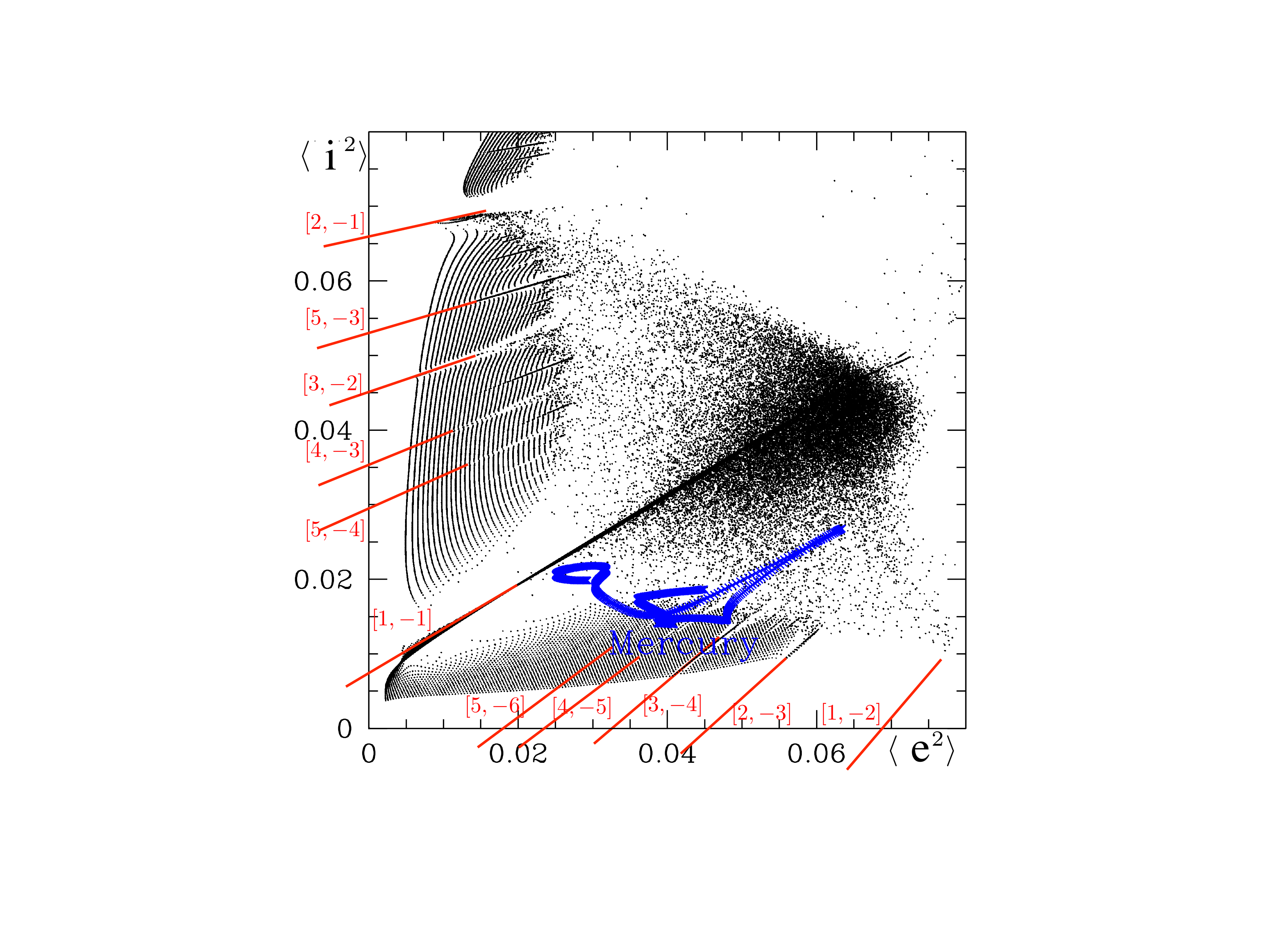}}
  \caption{The blue points show {\y the running average of} Mercury's
         $e^2$ and $i^2$ in an N-body simulation of the
    solar system {\y that lasts 600 Myr.  The
      black points  illustrate where chaos occurs in
      a highly simplified model that includes only the
      $g_5$ and $s_2$ forcing frequenices.  Those points
                denote the time-averaged results
        from simulations that are initialized on a regular grid;
             hence the
      regularly spaced black points denote non-chaotic orbits, and the
      irregularly spaced points denote chaotic ones.  The fact that
      Mercury's orbit lies within the chaotic zone shows that our
      highly simplified model is sufficient to account for Mercury's
      chaos.   Moreover,
        this model demonstrates that Mercury is perched on the
      threshold of chaos. The red lines denote various secular
      resonances that involve the $g_5$ and $s_2$ frequencies.
      Adapted from Fig. 5 of LW11; see that paper for more detail.}
  \label{fig:pphi}
 \label{fig:messy}
}
 \end{figure}

Figure \ref{fig:messy} compares Mercury's orbital evolution in an
N-body simulation of the solar system (blue points) with the
prediction based on a simplified model that includes only the $g_5$
and $s_2$ forcing terms.  Mercury's true orbit traces the boundary of
chaos as predicted by the model, illustrating that the model suffices
to explain Mercury's chaos.  More dramatically, it also illustrates
how Mercury is perched on the threshold of chaos.  We speculate below
as to how Mercury might have ended up in such a seemingly unlikely
state.

\section{Hot Jupiters}

The first batch of extra-solar planets that were discovered were ``hot
Jupiters'' \cite{MayorQueloz,marcybutler96}. It 
is now
 clear that
$\sim 1\%$ of solar-type stars are orbited by jovian giant planets
with periods of $\sim 3$ days. 
 In comparison with this  pile-up of hot Jupiters at  small
separation \cite{Gaudi,Butler,Cumming,Fressin}, 
there is a 
deficit of gas giants with periods between $10$ and $100$ days
(the ``period valley;'' \cite{UdryMayorSantos,Wittenmyer}) before
the number picks up and rises outward again 
(see reviews
\cite{Marcy,Udry}).

According to 
{\y conventional} theories of planet formation, hot Jupiters could not
have formed in situ, given the large stellar tidal field, high gas
temperature, and low disk mass to be found so close to the star.  It
is therefore commonly thought that these planets are formed beyond a
few AU and then are migrated inward.  Candidate migration scenarios
that have been proposed include protoplanetary disks
(e.g., \cite{Lin86,Lin96}), Kozai migration by binary or planetary
companions (e.g., \cite{WuMurray,FabryckyTremaine,Naoz}), scattering
with other planets in the system (e.g., \cite{FordRasio2}), and
secular chaos \cite{WuLithwick,Nagasawa}. A critical review of these
mechanisms is given in WL11.

Here we present a short description of the secular chaos
scenario. Moreover, now that the orbital axis (relative to the stellar
rotation axis) of some $60$ hot Jupiters has been measured, we
  compare
the observed distribution against 
{\y that produced in a  new suite of secular chaos simulations}.

\subsection{Simulations and Results}

Our fiducial planetary system is composed of three giant planets
($0.5, 1.0, 1.5 M_J$) that are well-spaced ($1,6,16$ AU) with
initially mild eccentricities and inclinations ($e =0.07$
to $0.3$, inclination $4.5$ to $20 \deg$, see Table 1 of
WL11).
    Such a configuration is possible for a system that
emerges out of a dissipative proto-planetary disk, as it avoids
short-term instabilities.  {\y 
   But disk-planet interactions are not yet well understood. It is also possible that 
     disks always damp planets onto nearly circular coplanar
  orbits,  in which case such $e$'s and $i$'s might arise from,
  e.g., planet scattering or collisions.  }

The angular momentum deficit is defined as 
(e.g., \cite{Laskar97,ogilvie} )
\begin{equation}
{\rm AMD} \equiv \sum_k m_k \sqrt{a_k} (1 - \sqrt{1 - e_k^2} \cos i_k)
\approx \sum_k {1\over 2} m_k \sqrt{a_k} (e_k^2 + i_k^2)\, ,
\label{eq:AMD}
\end{equation}
where the summation is over all planets. The AMD describes the deficit in
angular momentum relative to that of a coplanar, circular system.
When the AMD is not zero, secular interactions continuously modify the
planets' eccentricities and inclinations, preserving the total AMD
(since the total angular momentum is conserved, and secular
interactions do not modify the orbital energies). A system with a
higher AMD will interact more strongly, and above some critical AMD
value, the evolution is chaotic (e.g., Fig. \ref{fig:resoverlap}). In
order to produce a hot Jupiter by secular interactions (requring that
$e_1 \rightarrow 1$), the minimum AMD value is ${\rm AMD} = m_1
\sqrt{a_1}$. Our fiducial system has an AMD of $1.17 m_1 \sqrt{a_1}$.
This AMD value is also high enough for the system to be secularly
chaotic.

\begin{figure}
\includegraphics[width=0.35\textwidth,angle=-90,
trim=180 140 50 100]{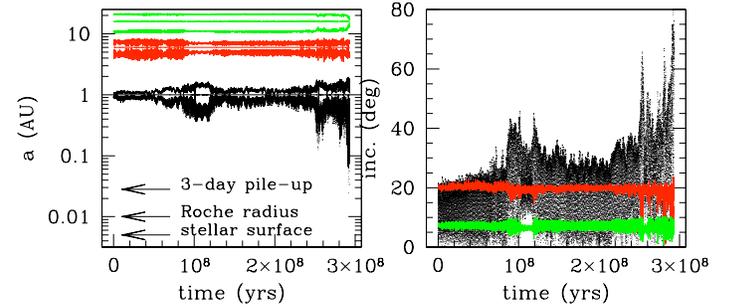}
\caption{
Formation of a hot Jupiter in our fiducial system, simulated using
the SWIFT code  \cite{levisonduncan} with tidal and GR effects  added. Left: radial excursions of the three
  planets (semimajor axis, periapse and apoapse) are shown as
  functions of time, with the various radii relevant for hot Jupiters
  marked by arrows; right: planet inclinations measured relative to
  the system's invariable plane. All planets initially have mildly
  eccentric and inclined orbits, but over a period of $300$ Myr so
  much of the angular momentum in the innermost planet can be removed
  that its eccentricity and inclination can diffusively reach order
  unity values.  The semimajor axes
    remain nearly constant until the end, a tell-tale sign that secular interactions dominate the
  dynamics.  At $\sim 300$ Myr, the pericenter of the inner planet
  reaches inward of a few stellar radii and tidal interaction with the
  central star kicks in.  Precessions by general relativity, by tidal
  and rotational quadrupoles, as well as tidal dissipation, prevent the
  pericenter from reaching inward of the Roche radius. As a result,
  the final hot Jupiter has a period of $\sim 3$ days. Adapted
from WL11.
  \label{textbook}
}
\end{figure}

In Figure \ref{textbook}, we observe that the three planets secularly
(and diffusively) transfer angular momentum (but not energy) for
almost 300 Myrs without major mishap, until the inner planet has
gradually acquired so much AMD that its eccentricity, starting at
$e_1=0.07$, has reached $e_1 = 0.985$.  This corresponds to a periapse
distance of order the Roche radius, $a(1-e)$
{\y $\sim$}$ 0.015$AU.  \footnote{\y  Since the
  strength of tidal damping  
   rises very rapidly with
  decreasing peripase,  the Roche radius
  \newline
  (inside of which the planet would be shredded)  roughly
    characterizes 
         the distance at which tides 
         \newline
         stall any further
    periapse decrease .  }
We then specify in our numerical simulation that tidal interaction
with the central star removes orbital energy from the planet {\y
  \cite{Hut}}. The orbit decays inward, until the planet is tidally
circularized at $\sim 2\times$ Roche radius {\y (because of angular
  momentum conservation)}. The inner planet is now captured into a
`hot Jupiter'.

Since the AMD is transferred to the inner planet to raise its
eccentricity, the outer two planets end up with reduced eccentricities
and mutual inclinations. They remain at large separations, waiting to
be probed   by techniques such as  radial velocity, astrometry or
gravitational lensing. By getting rid of 
 their inner companion
 the remaining planets organize themselves
 into a more stable configuration,
analogous to 
 what would happen in the inner Solar system  after the loss of Mercury.

\begin{figure}
\includegraphics[width=0.39\textwidth,angle=0,
trim=20 120 50 150,clip=true]{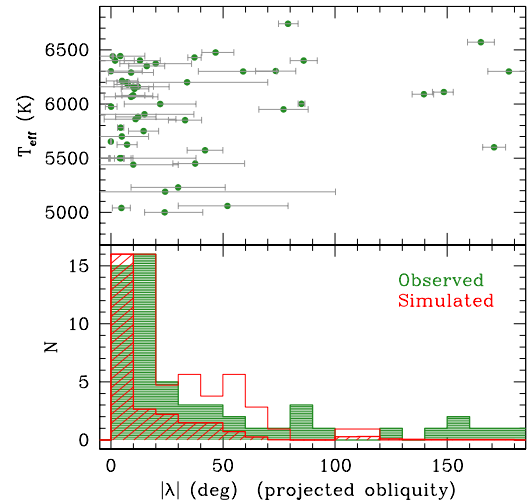}
\caption{The upper panel shows the sky-projected spin-orbit angles
  ({\y in} absolute value) for some $60$ hot Jupiters ($m\sin i \geq
  0.3 M_J$, data taken from exoplanets.org) as a function of host star
  effective temperatures. The lower panel shows the distribution of
  this projected angle (green shaded histogram). The {\y two red
    histograms show the angle distribution obtained from our simulations
    before projection (unshaded) and after projection (shaded).}
Secular chaos tends to produce
  prograde hot Jupiters, with the projected obliquities peaking at
  alignment, {\y and a significant tail extending to $\sim 50\deg$}.
  \label{RMangle}
}
\end{figure}

In addition to the showcase in Fig. \ref{textbook}, we have performed
{\y for this contribution} 100 simulations with AMD$= 1.5 m_1
\sqrt{a_1}$, $50\%$ over the minimum criterion.\footnote{ {\y
     {\y This AMD corresponds to
      $e, i\sim 0.3$.  Such $e$'s are typical of those
      measured for extrasolar
      \newline
      giant planets (not hot Jupiters). The
      origin of this AMD, however, is beyond the scope 
      of this
      review.} }}  The planets were initially at 3, 15 and $a_3$ AU,
with $a_3$ uniformly distributed between $30$ and $60$ AU.  We find
that roughly $60\%$ of these systems produce a hot Jupiter. Most of
these newly formed Jupiters have orbits that are prograde relative to
the original orbital plane, but some can be retrograde
(Fig. \ref{RMangle}{\y ---to be discussed in more detail
  below}). Moreover, the time for secular chaos to excite the orbital
eccentricities to tidal capture ranges from a few million years to a
hundred-million years. Raising the initial AMD leads to more efficient
hot Jupiter formation.

\subsection{Secular Chaos Confronting Observations}
In the following, we compare the predictions of secular chaos with
observations, highlighting the distribution of spin-orbit
angles.

There is a sharp inner cut-off to the 3-day pile-up of hot
Jupiters. They appear to avoid the region inward of {\it twice} the
Roche radius \cite{FordRasio}, where the Roche radius is the distance
within which a planet would be tidally shredded.  New data spanning
two orders of magnitude in planetary masses (and including planet
radius measurements) have strengthened this claim. There are only 5
known exceptions lying inward of twice the Roche radius, and the rest
mostly lie between twice and four times the Roche radius. 
 Mechanisms
that rely on eccentricity excitation, such as Kozai migration or
planet-planet scattering, naturally produce hot Jupiters that tend to
avoid the region inside of twice the Roche radius \cite{FordRasio}.
However, only Kozai migration and secular chaos  naturally 
produce a pile-up  
{\y just outside} twice the Roche radius, as the eccentricity rise in
these cases is gradual and planets are accumulated at the right
location.

Hot Jupiters appear to be less massive than more distant planets
\cite{Patzold,Zucker,2009ApJ...693.1084W}.  Among planets discovered
with the radial velocity method, close-in planets typically have
projected masses ($M\sin i$) less than twice Jupiter's mass.  But
numerous further out planets have $M\sin i>2M_J$
(Fig. 5 of \cite{Udry}).  This is expected in the context of secular
chaos (but not {\y the} Kozai mechanism). Since the minimum AMD to
produce a hot Jupiter rises with the planetary mass, we expect hot
Jupiters to be lower-mass than average.

Secular chaos predicts that hot Jupiters may have misaligned orbits
relative to the  invariable plane of the system. Here, we use the stellar
spin axis as the proxy for the latter, assuming that the stellar spin
is aligned with the proto-planetary disk in which the planets were
born.  The spin-orbit angle can be probed in cases where the hot
Jupiter transits  its star, via the Rossiter-McLaughlin
(R-M) effect (e.g., \cite{winn2005}).
The sky-projected value of the stellar obliquity has been
reported for some $60$ hot Jupiters (Fig. \ref{RMangle}).
While a majority of the hot Jupiters are aligned with the stellar
spin, a smattering of them (especially those around hotter
stars; \cite{winn10,Triaud,albrecht}) appear to have isotropic orbits.  The
observed distribution can be decomposed into one that peaks at alignment
and one that is isotropic \cite{Fabryckywinn}. 

Among the 60 hot Jupiters that formed in our set of 100 simulations,
the vast majority have prograde orbits (with only 2 retrograde ones).
This is because we initialized the simulations with  50\% more AMD
than the critical amount to form a hot Jupiter.  In that case, when a
sufficient amount of AMD has been transferred into the innermost
planet to increase its eccentricity to $\sim 1$, there is not much AMD
left to excite its inclination too.  In simulations with higher
initial AMD, more inclined hot Jupiters tend to be produced.  In our
100 simulations presented here, the spin-orbit angles are roughly
Gaussian distributed with a FWHM of $\sim 30\deg$. We project these
angles onto the sky, assuming that the systems are randomly
distributed relative to the line-of-sight {\y (Fig. \ref{RMangle})}.
The sky-projected obliquity (RM angle) is dominated by nearly aligned
planets, with RM angle for $30\%$ of the systems smaller than $
2\deg$, and $60\%$ of the system within $10\deg$. {\y But a
  significant tail, about $40\%$ of the systems, extends to $\sim
  50\deg$.  } This may explain the observed population of {\y prograde} planets, especially considering that
observational errorbars tend to broaden the distribution.  {\y We note
  that a} more coplanar mechanism like disk migration will
{\y likely} produce a {\y peak at alignment, but no  tail.}

Hot Jupiters also tend to be alone, { at least out to} a few AU.  From
radial velocity surveys, $\sim 30\%$ of planets are in multiple planet
systems (including ones with radial velocity trends, \cite{Butler}),
while only 5 hot Jupiters, {i.e. fewer than $10\%$ of hot Jupiters are
  known to have companions within a couple AU.}  This relative deficit
also shows up in the transit sample, where most attempts at detecting
transit timing variations caused by close companions {\y of hot
  Jupiters} \cite{HolmanMurray,Agol} have been unsuccessful
(e.g., \cite{Rabus,Csizmadia,Hrudkov,steffenpnas}). 
{\y That contrasts with the many TTV detections for other kinds
  of planets (e.g., \cite{Ford12,SteffenIII,FabryckyIV}).}  Such an
absence of close-by companions {\y to hot Jupiters} is consistent with
the picture that hot Jupiters had high eccentricities in the past.
Secular chaos also predicts that in systems with hot Jupiters, there
are  {\y at least two other} giant planets roaming at larger
distances. This is testable with ongoing long-term, high precision RV
monitoring.

Both secular chaos and Kozai migration predict that hot Jupiters are
migrated after the disk dispersal. So detection of such objects around
T Tauri stars can be used to falsify these theories.

\section{Organization of Planetary Systems by Secular Chaos}

The fact that the solar system is marginally stable might be hinting
at a deeper truth about how planetary systems are organized.  It seems
implausible that the solar system was so finely tuned at birth to
yield an instability time comparable to its age {\it today}.  Rather,
the solar system might have maintained a state of marginal stability
at all times \cite{Laskar96}.  In this scenario, the stability time was
shorter when the solar system was younger because there were more
planets then.  As the solar system aged, it lost planets to collision
or ejection.  Each loss lengthens the stability time because a more
widely spaced system is more stable. In this way, the solar system
would naturally maintain marginal stability.  The precarious state of
Mercury on the threshold of chaos (Fig. \ref{fig:messy}) might merely
be the last manifestation  of such a self-organizing process.
Similarly, hot Jupiters might be the most conspicuous evidence that
extrasolar systems also undergo such self-organization.

We suggest that secular chaos might be reponsible to a large extent
for organizing planetary systems.  In secular interactions, AMD is
conserved---one may think of AMD as the  free energy.  We
conjecture that secular chaos drives systems towards equipartition of
AMD, such that, on average, all secular modes have equal AMD.  That
would be consistent with the terrestrial planets, where the lightest
planets are the most excited ones.  Let us consider a possible
scenario for how planetary systems evolve 
(see also \cite{Laskar96}.
Initially, planets merge or are ejected until the
AMD is such that neighboring planets cannot collide in a state of
equipartition. The secular evolution on long timescales is then set by
fluctuations about equipartition---one planet (or more properly its
mode) happens to gain a sufficiently large portion of the AMD that it
merges with its neighbor, or is ejected, or approaches the star and
forms a hot Jupiter. After such an event, the AMD would decrease, and
the planetary system would be more stable than before. But on a longer
timescale fluctuations can once again lead to instability. Of course,
this scenario is speculative, and must be tested against simulations
and observations.  Fortunately, the hundreds of planetary systems
recently discovered provide a testbed for  such explorations.








\begin{acknowledgments}
 Y.L. acknowledges support from NSF grant
AST-1109776.
  YW acknowledges support by NSERC and the Ontario
government.
\end{acknowledgments}





\end{article}








\end{document}